\documentclass[12pt]{spieman}  
\usepackage{amsmath,amsfonts,amssymb}
\usepackage{graphicx}
\usepackage{setspace}
\usepackage{tocloft}
\usepackage{booktabs}
\usepackage{tablefootnote}
\usepackage{aas_macros}

\title{PRIMA: PRIMAger, a far-infrared hyperspectral and polarimetric instrument}

\author[a]{Laure Ciesla}
\author[b]{C. Darren Dowell}
\author[c]{Marc Sauvage}
\author[a]{Denis Burgarella}
\author[d,e]{Jochem Baselmans}
\author[f]{Matthieu B\'ethermin}
\author[b]{Jeffrey T. Booth}
\author[b]{Charles M. Bradford}
\author[g]{Florent Canourgues}
\author[h]{Ivan Charles}
\author[a]{Anne Costille}
\author[i]{Thomas Essinger-Hileman}
\author[d]{Lorenza Ferrari}
\author[a]{Johan Floriot}
\author[b]{Marc Foote}
\author[i]{Jason Glenn}
\author[b]{Renaud Goullioud}
\author[j]{Matt Griffin}
\author[k]{Oliver Krause}
\author[d]{Willem Jellema}
\author[b]{Elizabeth Luthman}
\author[a]{Laurent Martin}
\author[b]{Margaret Meixner}
\author[a]{Tony Pamplona}
\author[b]{Klaus M. Pontoppidan}
\author[l]{Alexandra Pope}
\author[h]{Thomas Prouv\'e}
\author[b]{Jennifer Rocca}
\author[i,m]{Johannes Staguhn}
\author[j]{Carole Tucker}

\affil[a]{Aix Marseille Univ, CNRS, CNES, LAM, Marseille, France}
\affil[b]{Jet Propulsion Laboratory, California Institute of Technology, 4800 Oak Grove Drive, Pasadena, CA, 91109, USA}
\affil[c]{Universit\'e Paris-Saclay, Universit\'e Paris Cit\'e, CEA, CNRS, AIM, 91191, Gif-sur-Yvette, France}
\affil[d]{Netherlands Institute for Space Research (SRON), Niels Bohrweg 4, Leiden 2333 CA, the Netherlands}
\affil[e]{Department of Microelectronics, Delft University of Technology, Mekelweg 4, Delft 2628 CD, the Netherlands}
\affil[f]{Observatoire Astronomique de Strasbourg, Universit\'e de Strasbourg, CNRS UMR 7550, 11 rue de l’Universit\'e, 67000 Strasbourg, France}
\affil[g]{center National d’Etudes Spatiales – center spatial de Toulouse, avenue Edouard Belin, 31401 Toulouse Cedex 9, France}
\affil[h]{Univ. Grenoble Alpes, CEA, IRIG, DSBT, France}
\affil[i]{NASA Goddard Space Flight Center, Code 665, Greenbelt, Maryland, United States}
\affil[j]{School of Physics and Astronomy, Cardiff University, Cardiff CA24 3AA, UK}
\affil[k]{Max Planck Institute for Astronomy, Heidelberg, Germany}
\affil[l]{Department of Astronomy, University of Massachusetts, Amherst, MA, 01003, USA}
\affil[m]{Physics\&Astronomy, Johns Hopkins University, 3400 N. Charles St, Baltimore, MD 21218, USA}

\cftpagenumbersoff{figure}
\cftpagenumbersoff{table} 
\begin{document} 
\maketitle

\begin{abstract}
The PRobe far-Infrared Mission for Astrophysics (PRIMA) is an infrared observatory for the next decade, currently in Phase A, with a 1.8\,m telescope actively cooled to 4.5\,K.
On board, an infrared camera, PRIMAger, equipped with ultra-sensitive kinetic inductance detector (KID) arrays, will provide observers with coverage of mid-infrared to far-infrared wavelengths from 24 to 264\,$\mu$m. 
PRIMAger will offer two imaging modes: the Hyperspectral mode will cover the 24-84\,$\mu$m wavelength range with a spectral resolution R$\geq$8, while the Polarimetric mode will provide polarimetric imaging in 4 broad bands, from 80 to 264\,$\mu$m. 
These observational capabilities have been tailored to answer fundamental astrophysical questions such as black hole and star-formation co-evolution in galaxies, the evolution of small dust grains over a wide range of redshifts, and the effects of interstellar magnetic fields in various environments, as well as to open a vast discovery space with versatile photometric and polarimetric capabilities.
PRIMAger is being developed by an international collaboration bringing together French institutes (Laboratoire d’Astrophysique de Marseille and CEA) through the center National d’Etudes Spatiales (CNES, France), the Netherlands Institute for Space Reseach (SRON, Netherlands), and the Cardiff University (UK) in Europe, as well as the Jet Propulsion Laboratory (JPL) and Goddard Space Flight Center (GSFC) in the USA.
\end{abstract}


{\noindent \footnotesize\textbf{*}Laure Ciesla,  \linkable{laure.ciesla@lam.fr} }

\begin{spacing}{1}   
\section{Introduction}
\label{sec:intro}  

Although major observatories in space and on the ground are currently observing in the near- to mid-infrared (\textit{James Webb} Space Telescope) and sub-millimeter (ALMA) wavelength ranges, the far-infrared (FIR) domain (30-300\,$\mu$m) remains inaccessible since the ends of the \textit{Herschel} Space Observatory in 2013 and SOFIA, the Stratospheric Observatory for Infrared Astronomy, in 2022, leaving the community blind to half of the luminous content of the universe. With this huge wavelength gap, myriads of sources deeply embedded within dust, and the coldest objects of the Solar System, are out of reach. This spectral window is nevertheless essential to our understanding of the first stages of galaxy evolution, the mechanisms of star formation, the evolution of dust, and the study of the origins of planetary systems and of our own Solar System.  While X-rays reveal energetic sources obscured by modest columns of dust and complement the FIR in characterizing those environments, most of the gas mass is relatively cold and can be traced by dust which emits almost all of the luminosity in the FIR.

Following the recommendations of the Astro2020 Decadal\cite{Astro2020} report, NASA issued a call for proposals to develop an Astrophysics Probe Explorer\cite{apexcall} (APEX mission), either in the X-ray or FIR domain, for a launch in 2031. PRIMA (PRobe far-Infrared Mission for Astrophysics\cite{Glenn25}) responded to the call with an observatory concept designed to address timely and fundamental questions about the growth of galaxies and the development of stellar systems and their constituents. It will observe the build-up of heavy elements, dust, stars, and black holes in galaxies, and trace the masses, and water content and mineralogy of protoplanetary disks to probe the growth of solar systems. The majority of the observing time ($>$70\%) will be devoted to General Observer (GO) programs, while focused PI programs will address key science with rapid release to support vibrant Guest Investigations (GI) using archival data.  
PRIMA has two instruments. 
FIRESS\cite{Bradford25}, a multi-purpose spectrometer, covers the 24 to 235\,$\mu$m range with four slit-fed grating spectrometer modules providing resolving power ($R$) between 85 and 130, and high resolution mode with R$\sim$2000-20000 thanks to a Fourier-transform spectrometer module (FTM). PRIMAger, is a far-infrared imager and the subject of this paper. These two instruments are highly complementary, with FIRESS sampling gas content and properties and PRIMAger probing dust and macro-molecules. 

PRIMAger is designed to address multiple key goals from the 2020 Decadal survey, sharpened by input from the international community through interactive workshops and example science cases.
Among 76 GO science cases envisioned for PRIMA\cite{Moullet23}, 35\% require only PRIMAger observations, and another third make use of both FIRESS and PRIMAger.
This significant interest from the community highlights the need for an imager operating in the FIR domain. The key features of this instrument are:
\begin{itemize}
    \item{mapping speed, angular resolution, and spectral sampling sufficient to generate large FIR-selected samples of distinct galaxies when the Universe was 25\% of its current age and at the peak of its star-forming activity, and map star-forming regions of the Milky Way;}
    \item{spectrophotometric wavelength coverage and resolution sufficient to detect Polycyclic Aromatic Hydrocarbon (PAH) features at the same period of the Universe history, but also to provide time domain surveys of star-formation in the Milky Way.}
    \item{sensitivity to the fractional linear polarization of dust emission in multiple FIR bands in order to map the magnetic structure in molecular clouds and star-forming cores in the Milky Way and the large-scale magnetic field in nearby galaxies, and also to probe the nature of the interstellar dust grains.}
\end{itemize}

Section~\ref{sec-approach} gives an overview of the PRIMAger design which incorporates the key features above.
The two focal planes of PRIMAger are described in Sections~\ref{sec-phi} and \ref{sec-ppi}, while the observation mode is explained in Section~\ref{sec-obsmod}. Section~\ref{sec-detector} provides a description of PRIMAger's detectors. The overall mechanical, thermal, and optical designs are detailed in Section~\ref{sec-design}. Estimated instrument sensitivities are given in Section~\ref{sec-sens} and a summary of PRIMAger's capabilities is given in Section~\ref{sec-sum}.

\section{Functional approach\label{sec-approach}}

PRIMAger is a FIR camera employing two focal planes of kinetic inductance detectors (KIDs) that observe simultaneously to cover the 24 to 264\,$\mu$m range where the bands are defined by quasi-optical filters and the KIDs themselves act as broadband detectors.  
The hyperspectral focal plane (referred to as PHI for PRIMAger Hyperspectral Imager), composed of two arrays of lens-absorber-coupled hybrid KIDs, covers the 24-84\,$\mu$m range with a spectral resolution of $R = \frac{\lambda}{\Delta\lambda} \geq 8$.  Linearly variable filters (LVFs) placed above the KID arrays provide the $R \approx 8$ passbands as well as a gradient of average transmitted wavelength along the long axis of the detector array.
The polarimetric focal plane (referred to as PPI for PRIMAger Polarimetric Imager), covering the 80-264\,$\mu$m range with four filters ($R\approx 4$), has lens-antenna-coupled hybrid KIDs.
Each pixel is sensitive to one out of three angles of linear polarization, and together they measure the $I$, $Q$, and $U$ Stokes parameters.

\begin{figure}[ht] 
 	\includegraphics[width=\textwidth]{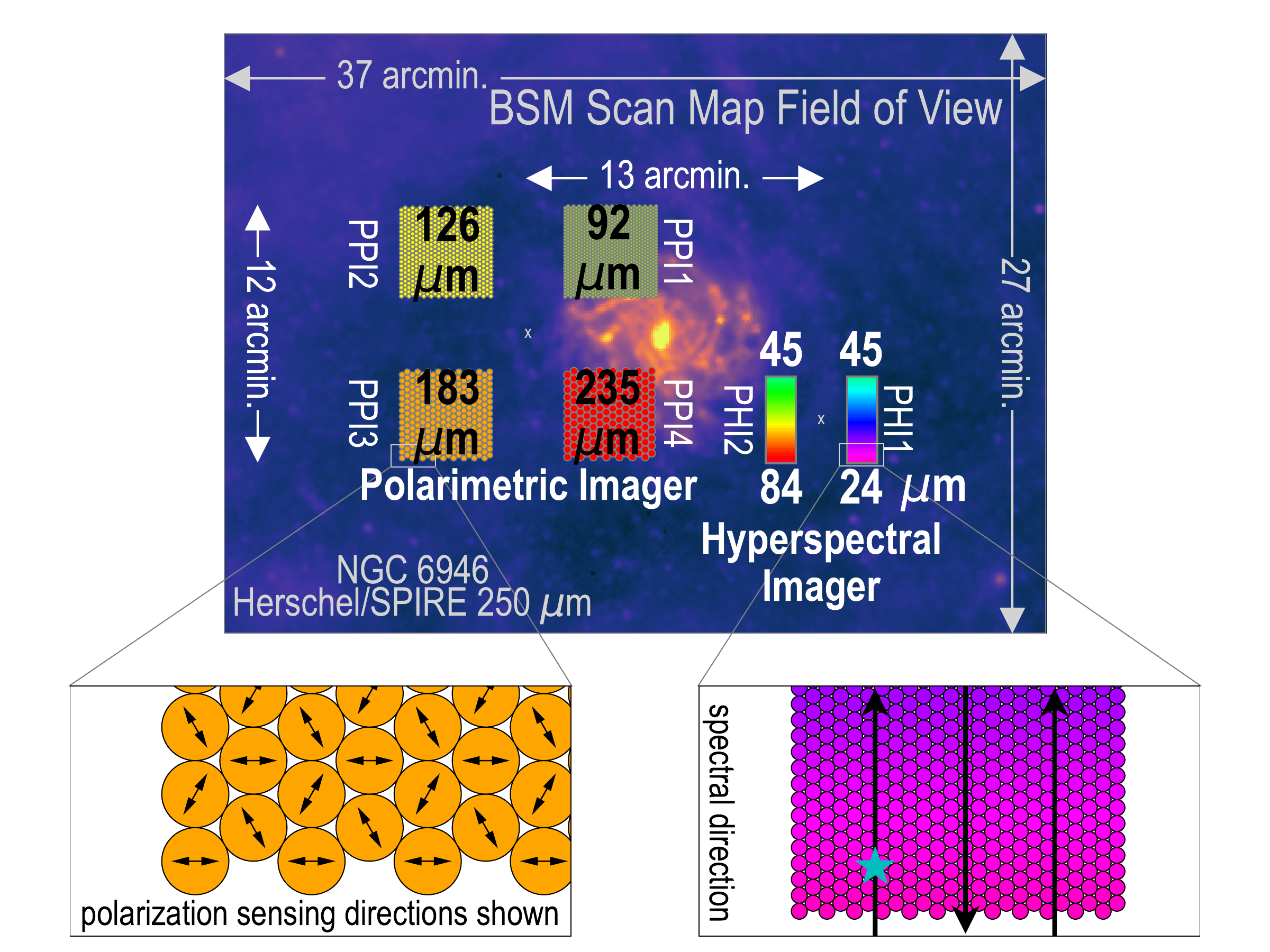}
  	\caption{\label{fig:fov} Field of view covered by PRIMAger. The projection of the PHI (PHI1 and PHI2) and PPI (PP1, PP2, PP3, and PPI4) focal planes on the sky are shown. From edge to edge, they are separated by $4.7^\prime$. The PRIMAger sub-fields can be moved within an area represented by the entire rectangle cut out from the \textit{Herschel}/SPIRE image of NGC\,6946 using the beam steering mirror. The left sub-panel show how each pixel of the PPI detector arrays is sensitive to one polarization angle. 
   The right sub-panel depicts how each PHI pixel bandpass depends on location in the array, with a gradient in central wavelength along the long axis thanks to the Linear Variable Filter.
   The black arrows show example directions of scanning a source to generate a spectrum.}
\end{figure}

The hyperspectral and polarimetric focal planes share the PRIMAger field of view and are separated by $4.7^\prime$ projected onto the sky, as shown in Fig.~\ref{fig:fov}.
A beam steering mirror (BSM) allows rapid, two-dimensional scanning of the instrument field of view within the $37^\prime \times 27^\prime$ telescope field of view. Larger areas will be mapped by combining this beam steering with scanning motion of the entire observatory. While each detector images a different part of the PRIMAger field of view, the scanning strategy ensures that a common sky area can be mapped by multiple detectors.

\subsection{PRIMA Hyperspectral Imager \label{sec-phi}}

\begin{figure}[ht] 
 	\includegraphics[width=\textwidth]{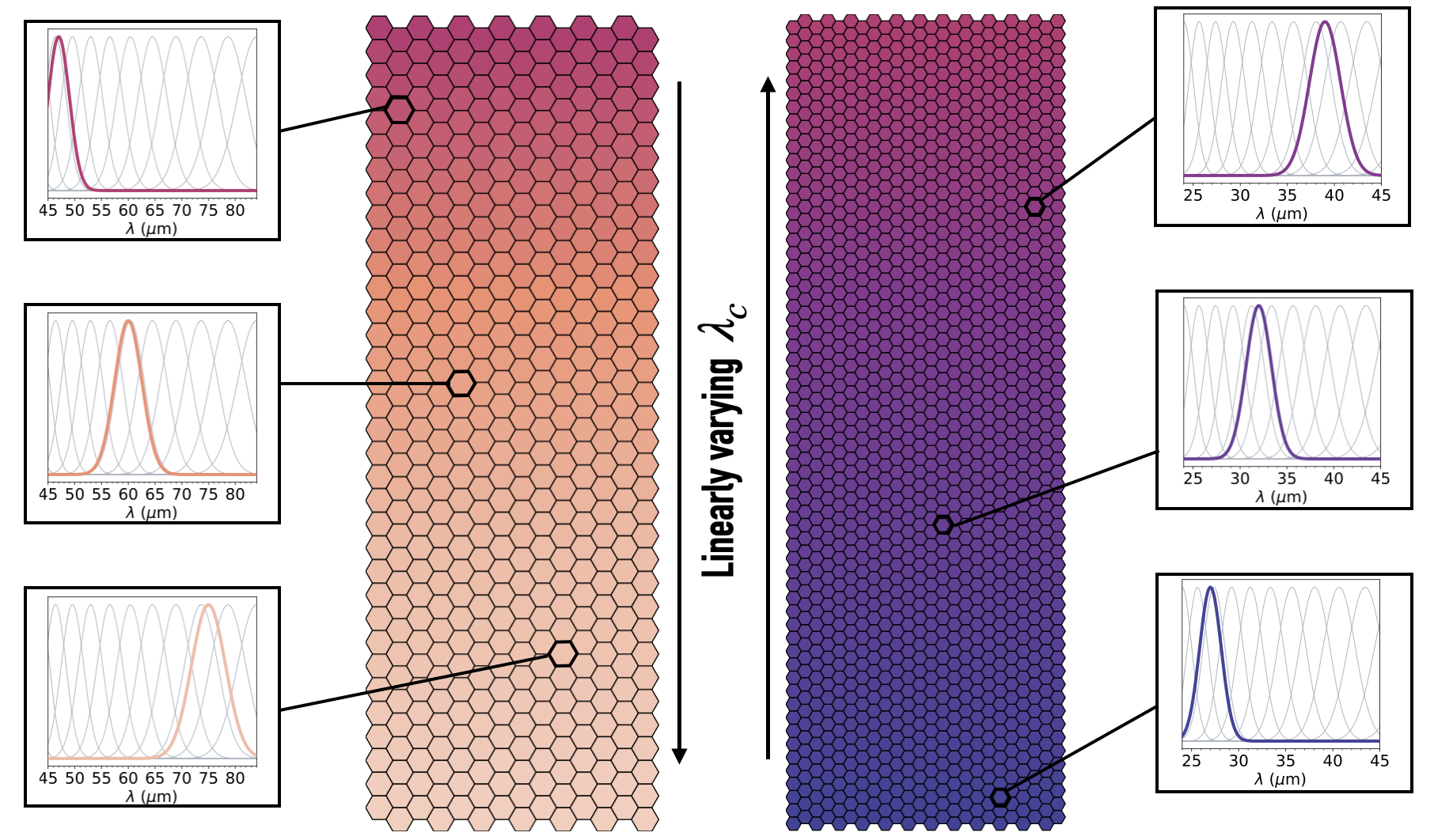}
      	\caption{\label{fig:phi_ill} Illustration of PHI detector arrays. PHI1 (24-45\,$\mu$m) is on the right while PHI2 (45-84\,$\mu$m) is on the left.  Each pixel, represented by black-edged hexagons, observes in an $R \approx 8$ band while the central wavelength depends on its position on the detector plane.}
\end{figure}

The PRIMAger Hyperspectral Imager (PHI) covers the 24 to 84\,$\mu$m wavelength range with two sub-bands (PHI1 and PHI2).
Spectral coverage is produced instantaneously as a spatial pattern on the sky, as the LVF is in a fixed position relative to the detectors.  To achieve spectral coverage of a particular source, the BSM or observatory scans the source along the long dimension of the detector arrays, through the wavelength interval $[24,45]$\,${\mu}$m or $[45,84]$\,${\mu}$m, for the PHI1 or PHI2 arrays, respectively (Fig.~\ref{fig:phi_ill}). The two arrays are fed with an f/21 optical path, which, in combination with the design detector pixel pitches, produce f$\lambda$ instantaneous spatial sampling of the focal plane (corresponding to angle lambda/D of the sky).  This is coarser than Nyquist sampling (f$\lambda$/2), therefore therefore achieving full angular resolution requires use of a scanning strategy (see Sect.~\ref{sec-obsmod}).

\subsection{PRIMA Polarimetric Imager \label{sec-ppi}}

The PRIMAger Polarimetric Imager (PPI) simultaneously observes with four detector arrays through monochromatic, broad-band filters ($R \approx 4$) with central wavelengths of 92, 126, 183, and 235\,$\mu$m.  The detectors are sensitive to linear polarization along three orientations as illustrated in the bottom left panel of Fig.~\ref{fig:fov}.  The detector optics and spacing are designed to deliver high mapping speed with a relatively small number of detectors and beam sizes near the diffraction limit.  All of the detectors are fed with a common \textit{f}/12 optical path, so the detector physical pitch scales approximately with wavelength. Here as well the instantaneous spatial sampling is $F\lambda$, and therefore the scanning strategy has to achieve the full angular resolution.

PPI polarimetry requires a scan pattern that observes each map pixel with the three polarization orientations.  The process of recovering maps of the Stokes parameters $(I, Q, U)$ from PPI data has been studied in detail \cite{Dowell2024}.  The method is a straightforward extension of mapping in total intensity\cite{Waskett07,Dowell2010,Roussel2013}, in that two-dimensional crossing scans allow the relative detector `baseline' signals to be determined and removed from the map, except for an unknown spatially-constant zero level.  Polarimetry requires the solution of three coupled linear equations for each map pixel, rather than the single equation for total intensity only.  The system of equations effectively performs the signal differencing inherent to polarization, and since FIR polarization signals tend to be weak ($\lesssim 20$\% of total intensity), systematic errors must be controlled to limit residual artifacts in the map.  This is accomplished with a combination of uniformity in the design and manufacture and correction in the science data pipeline.

\begin{table*}
    \centering
    \caption{Design characteristics of the two PRIMAger focal planes, PHI and PPI.}
    \begin{tabular}{l c c c c c c}
    \toprule
    \textbf{Parameter}      & \multicolumn{2}{c}{PRIMA Hyperspectral Imager}  & \multicolumn{4}{c}{PRIMA Polarimetry Imager}\\
    \midrule
                            &   PHI1    &   PHI2    &  PPI1 & PPI2 & PPI3 & PPI4 \\
    \toprule
    Central wavelength ($\mu$m) & 24-45 &  45-84 & 92 & 126 & 183 & 235  \\
    \midrule
    Spectral resolving power     & 8 & 8  & 4 & 4 & 4 & 4  \\
    \midrule
    Polarimetry             & - & -  & Yes & Yes & Yes & Yes  \\
    \midrule 
    Focal ratio (f$\lambda$)             & 21 & 21  & 12 & 12 & 12 & 12  \\
    \midrule  
    Band center FWHM (")       &  4.7     & 8.7    & 10.9 & 14.9 & 21.7 & 27.9  \\
    Pixel size      (")         &  3.8     & 6.8    & 8.6 & 11.6 & 16.1 & 20.2  \\
    \midrule
    Pixel count              &   61$\times$24     & 34$\times$14     &  34$\times$29 & 25$\times$21   & 18$\times$15  &  14$\times$12  \\
    \midrule
    Field of view           &   3.9'$\times$1.3' & 3.9'$\times$1.4' &  4.2'$\times$4.2' & 4.2'$\times$4.2' & 4.2'$\times$4.2' & 4.2'$\times$4.1'     \\
    \bottomrule 
    \label{table:characteristics}
    \end{tabular}
\end{table*}

\section{Observing Modes \label{sec-obsmod}}

PRIMAger is designed to map areas significantly larger than the intrinsic field-of-view (FoV) of its individual arrays ($\sim 4^\prime \times 4^\prime$).  
This is achieved by implementing a spacecraft scanning pattern that covers large areas of the sky, designed to achieve a homogeneous coverage pattern with as little variation in depth as possible.
In addition, PRIMA carries a beam steering mirror (BSM) that can either be used to position the Line-of-Sight (LoS) within the telescope FoV in any direction without moving the spacecraft, or to provide further spatial modulation to a scan made by the spacecraft. 

PHI and PPI detectors have particular features that place further constraint on the observing mode.
Regarding PHI, all rows of the array must intercept the source of interest to obtain its full hyperspectral capability, thus the scanning pattern must ensure that the source crosses the full array along its long dimension. For PPI, the polarization information requires that the source of interest is observed by all three different type of pixels, but as these are always in direct vicinity of one another, this is realized by almost any scanning pattern.
For both PHI and PPI, scanning is mandatory to reconstruct the information, and there will be no staring or snapshot mode.

In this section, we first present the BSM and then describe the large map and small map scanning modes.

\subsection{The beam steering mirror}
\begin{figure*}[ht] 
        \centering
 	\includegraphics[width=3in]{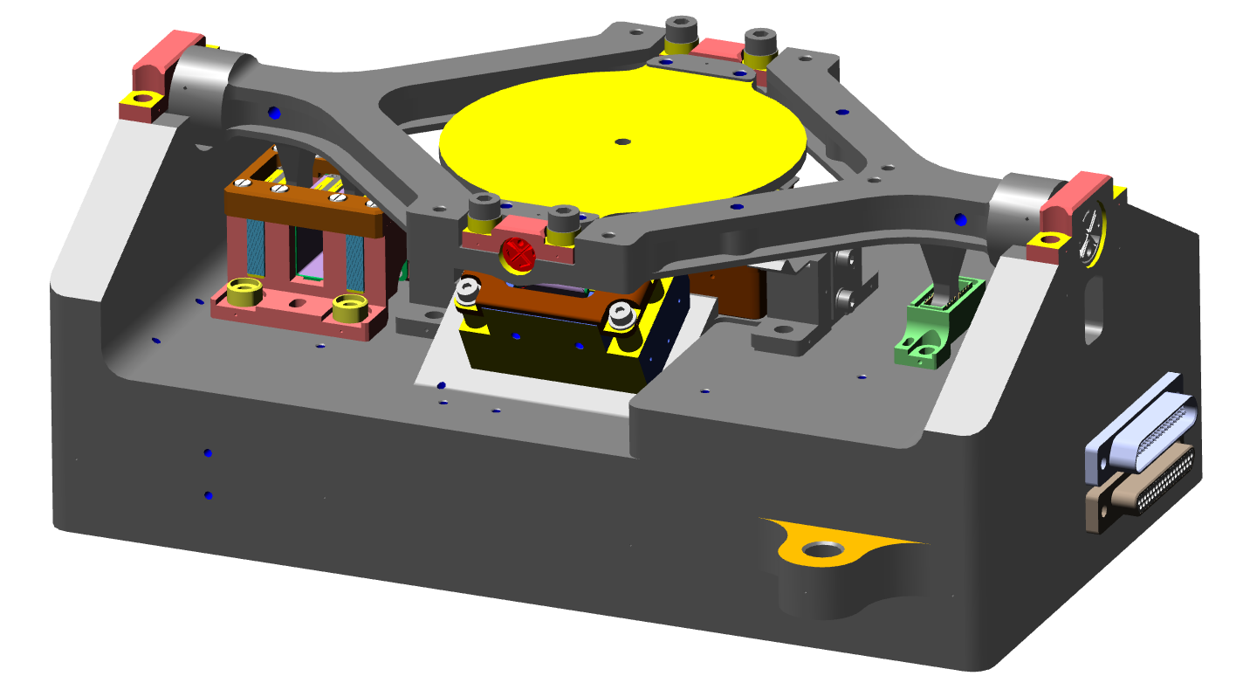}
  	\caption{\label{fig:optics_bsm} The PRIMAger two-axis BSM uses \textit{Herschel} PACS heritage and allow up to $\pm$10' steering on the sky, enabling PRIMAgers’s mapping modes.}
\end{figure*}

The BSM (Fig.~\ref{fig:optics_bsm}), located at the entrance pupil of PRIMAger instrument, enables signal modulation and flexible mapping modes. The 60\,mm aperture cryogenic mirror mechanism builds on \textit{Herschel} Photodetector Array Camera \& Spectrometer (PACS) heritage\cite{Krause06}. The system is optimized for low and stable cryogenic power dissipation, smaller than 3\,mW, at the BSM operating temperature of 4.5\,K. Requirements and capabilities are two-axis motions covering up to $\pm$10' on the sky with positional accuracy of better than 0.33" RMS, and arbitrary patterns including for example Lissajous, boustrophedon, and triangular ones. The motion speed is up to 30'/sec and maximum angular accelerations up to 20°/sec$^2$ on sky during transition times of 50\,ms for scan direction reversals. The mirror is located on a gimbal stage and suspended in both axes by two pairs of flex pivots. These monolithic pivots are fabricated with beryllium copper and provide enough lifetime for the large number of BSM motion cycles. The sensor system for closed-loop control consists of magneto-resistive field plates that are biased by permanent magnets located on the backside of the mirror. The BSM is actuated by two pairs of DC motors based on the \textit{Herschel} PACS chopper. They consist of stationary drive coils made of high-purity aluminum and NdFeB permanent magnets located on the mirror and gimbal assembly. The BSM is controlled via a dedicated electronics using specific waveforms for a feed-forward commanding of the respective motion patterns. A  closed-loop control with the BSM sensors allows to meet the stringent positional accuracy requirements. The BSM is provided by the Max Planck Institute for Astronomy (MPIA) to the PRIMA project.

\subsection{Large maps: simple scanning with the spacecraft}
\label{sec-largemaps}
The simplest observing mode is to scan the sky along a quasi-rectangular grid, in a so-called ``boustrophedon'' mode or raster scan. 
In this mode the spacecraft performs long regular slews in one direction, called scan legs, followed by a small step in the perpendicular direction, followed by another scan leg parallel to the first one, reversing the scan direction. 
The length of a scan leg will typically range between some tens of arcminutes (e.g., when mapping individual objects and maximizing the spatial sampling density) to a few degrees (e.g., when mapping complete interstellar clouds, sections of the Galactic plane, or extragalactic fields). The spacing between consecutive scan legs sets the pattern of coverage depth in the resulting map, and is usually adjusted to provide a central mapped area with as homogeneous a depth as possible.

To mitigate the effects of 1/\textit{f} noise, it is advantageous to perform scans at multiple angles in an observation.
Thus for PPI we can use the {\em Herschel} approach and systematically combine two scans performed with approximately perpendicular scanning directions. For PHI this will not be possible as the PHI array lacks symetry in the spectral sampling direction, thus 1/\textit{f} mitigation will require the use of the BSM (see below).

Further characterization of the detector behavior will allow the optimization of this observing mode. The key drivers here are: the low-frequency noise ``knee'' frequency which constrains the signal modulation frequency, setting a lower limit on the scanning speed; the detector's time constant (or instead the telemetered sampling rate), which sets an upper limit on the scanning speed to avoid beam smearing; and the spacecraft maneuvering agility, which determines an observing time cost function for any scanning pattern. Generally speaking, as any spacecraft maneuver is costly, preferred configurations will have long scan legs, with minimal overlaps between back and forth scans.

The scanning mode is also constrained by the fact that it needs to reconstruct the full spatial sampling of the observed scene: as mentioned before, detector arrays on PRIMAger provide only $\sim F\lambda$ spatial sampling, therefore the scanning direction defines how oversampled the resulting map can be with respect to the intrinsic array spatial sampling. Considering the hexagonal arrangement of pixels in our focal plane, a preferred scan direction is 10.89 degrees from vertical (as defined from Fig.~\ref{fig:fov}), as individual pixel trajectories on the sky are then evenly spaced and provide a spatial sampling $\sim F\lambda/3$ on the sky. Similar considerations were used by the {\em Herschel}/SPIRE team to define the scanning directions for its large map mode\cite{Griffin10}.

\subsection{Small maps: scanning with the beam steering mirror}

As the target area becomes smaller, observing it with spacecraft scanning becomes increasingly inefficient, as more time is spent turning the spacecraft around than scanning through the target. Therefore for fields that are smaller than the $37'\times27'$ field of view delivered to PRIMAger by the telescope, we will scan the scene with the Beam Steering Mirror (BSM) only. The BSM is extremely agile and allows any kind of 2D trajectory within the field of view, as well as a very large range of angular speeds. The constraint set by modulating the signal above the noise knee frequency will be easily met by the BSM motion, and as above, the angular speed of the BSM will be adjusted to avoid beam-smearing due to the response time of the detectors. As the BSM is able to perform any 2D pattern (e.g. Lissajous or pong-like), detector spatial sampling needs will be easily met.

\subsection{Optimized mapping combining spacecraft and beam steering mirror mapping}

Combining spacecraft and BSM scanning in mapping mode can prove optimal in certain situations (either astrophysically motivated or linked to the detector properties):
For scenes that are larger than the telescope field of view yet not reaching the degree scale (i.e. when the cost of spacecraft motion is significant), we can minimize the required number of scan legs with the spacecraft by performing rapid perpendicular motions with the BSM. These motions may even provide the diversity of scanning angles used to beat $1/f$ noise, dispensing with the need for a perpendicular scan map.
Similarly, if the scanning speeds allowed by the spacecraft are not appropriate to modulate the signal above the $1/f$ knee frequency, rapid patterns with the BSM can provide these modulations.
It is in fact anticipated that a combination of the BSM scanning in a quasi-periodic pattern, such as Lissajous, with the telescope scanning slowly over a large area on the sky, will be the optimal observing mode for large surveys. Furthermore, the BSM scanning will provide the spatial sampling so that spacecraft scanning directions will not be constrained by the detector geometry as above.
Therefore mapping with PHI will always require the BSM.

\section{Sensitivity Requirements and Predictions\label{sec-sens}}

The PRIMAger team maintains a model for instrument performance based on astrophysical backgrounds, expected characteristics of the optics and detectors, and observing modes.  This model has influenced the scope of the PRIMA PI science objectives\cite{Glenn25,Burgarella25} and led to the establishment of the baseline survey science requirements that are given in Table~\ref{table:characteristics}.  In all cases, the expected instrument performance is significantly better than the baseline survey requirements, providing large margin for achieving the science objectives.

A key objective of PRIMA is to generate samples of distant galaxies from unbiased far-IR surveys and to study the evolution of the star formation and supermassive black hole accretion components with redshift.  Wide (larger area) and deep (smaller area) surveys have been designed with commensurate demands on sensitivity.  For the deep survey covering 1 degree$^2$ on the sky, the science-required point-source statistical uncertainty $5\nu \sigma(F_\nu)$ has been set at $1.2\times 10^{-17}$\,W/m$^2$, across the PRIMAger wavelength range, to be accomplished in 1500 hours.  (The factor of 5 corresponds to the target minimum signal-to-noise for the galaxy sample.)  This sets the point-source, total flux density requirements in Table~\ref{table:characteristics}:  $\nu \sigma(F_\nu)$ is divided by the bandpass center frequency $\nu$, and then scaled by 1/$\sqrt{\rm time}$.

For PPI, the derived point-source sensitivity requirement applies to each of the four monochromatic bands.  For PHI, which has a gradient in response wavelength, the requirement applies to an aggregation of detectors covering a 10\% range in wavelength, corresponding to the bandwidth of an $R = 8$ LVF.  The PPI sensitivity for polarized flux density is generated by multiplying the sensitivity to total flux density by $\sqrt{2}$ \cite{Dowell2024}.

The PPI surface brightness sensitivity requirements are driven by the science objective to map the polarized intensity of resolved galaxies and to study the nature of the dust by measuring the spectral evolution of the polarization fraction.  Consideration of the known surface brightness of galaxies (in total intensity) and estimated fractional polarization has led to a sensitivity requirement from 92 to 235 $\mu$m of $5\sigma(P_\nu) \leq $ 0.030 MJy/sr\cite{Dowell2024}, where $P_\nu$ is polarized surface brightness.  This sensitivity is to be achieved in 2 hours, over a 10 arcmin$^2$ area, and for an effective beam area (from post-processing) matching the 235 $\mu$m diffraction beam area, approximately 600 arcsec$^2$.  To generate the values for all PPI bands in Table~\ref{table:characteristics}, the polarized surface brightness requirement is scaled by 1/$\sqrt{\rm time}$ and 1/$\sqrt{\rm survey\ area}$, as well as by the inverse ratio of the band central wavelength to 235 $\mu$m to provide the sensitivity requirement at the nominal diffraction beam area.  The total intensity surface brightness sensitivity is set to 1/$\sqrt{2}\times$ the polarized intensity surface brightness sensitivity.

The PRIMA PI science program does not place requirements directly on the PHI surface brightness sensitivity.  For the purpose of completing Table~\ref{table:characteristics} and informing potential GO investigations, the point-source flux density requirements have been converted to surface brightness requirements using the beam area and point source coupling fraction (averaged over position within the detector array) from the instrument model.

The baseline survey requirements and other instrument performance requirements that have been established govern the design of PRIMA and PRIMAger going forward, encompassing aspects such as telescope wavefront error, transmission through the optical elements, detector noise, detector pixel count, and stray light.  Following standard practice, science and engineering margins have been set up between the science requirements and Current Best Estimates (CBE) provided by the instrument model.  As PRIMAger matures from design concept to tested instrument, the CBE will be frequently updated to reflect new results from design changes and ground testing.  As PRIMA nears launch, a set of sensitivity predictions will be published, expected to fall within the range of the science requirements and the present-day CBE.

Sensitivities in Table~\ref{table:characteristics} correspond to observations made in the {\em Large maps} mode (Sec.~\ref{sec-largemaps}), hence the choice of a reference observation of 1 square degree in 1\,hr. This mode involves significant overheads imposed by spacecraft maneuvering.
Considering the much smaller angular sizes of the PHI with respect to the PPI arrays, some efficiency gains exist for observations using only PPI.

An important component affecting the achieved sensitivity is spatial confusion (the contribution of undetected source to background fluctuations). As estimating the impact of confusion is heavily dependent on the science objective (through the data processing strategy) and on source count models, this is investigated specifically in independent papers\cite{Bethermin24,Donnellan24}.

To conclude this section, we remark, for comparison, that  \textit{Herschel}/PACS at 100\,$\mu$m reached a 5$\sigma$ instrumental noise, for a 15'$\times$10' map taking 30\,h, of 5\,mJy\cite{Poglitsch10}. At the science requirement level, PRIMAger is already 10 to 25 times deeper for fixed integration time, or up to 200 times faster for fixed target sensitivity, and current best estimates suggest that these numbers are lower limits to the gain PRIMAger represents compared to \textit{Herschel}/PACS.

\begin{table*}[!htbp]
    \centering
    \caption{Baseline Survey Sensitivity Requirements for PRIMAger.}
    \begin{tabular}{l c c c c c c}
    \toprule
    \textbf{Source type}      & \multicolumn{2}{c}{PRIMA Hyperspectral Imager}  & \multicolumn{4}{c}{PRIMA Polarimetric Imager}\\
    \midrule
                            &   PHI1    &   PHI2    &  PPI1 & PPI2 & PPI3 & PPI4 \\
                            &   24 -- 45 $\mu$m   &   45 -- 84 $\mu$m   &  92 & 126 & 183 & 235 $\mu$m \\
    \toprule
    Point Src. Flux Dens. (total, $F_\nu$; mJy) & 1.18 -- 2.2 & 2.2 -- 4.1 & 1.77 & 2.56 & 3.39 & 4.59 \\
    \midrule
    Point Src. Flux Dens. (polarized, p$F_\nu$; mJy) & -- & -- & 2.50 & 3.62 & 4.65 & 6.49 \\
    \midrule
    Surf. Bright. (total, $I_\nu$; MJy/sr) & 1.64 -- 0.66 & 0.74 -- 0.58 & 0.46 & 0.34 & 0.25 & 0.18 \\
    \midrule  
    Surf. Bright. (polarized, $P_\nu$; MJy/sr) & -- & -- & 0.65 & 0.47 & 0.35 & 0.25 \\
    \bottomrule 
    \label{table:characteristics}
    \end{tabular}
    {\raggedright The values above correspond to the 5$\sigma$ background-subtracted flux density limit in a 1 degree$^2$ map observed for a total duration of 10\,h (overheads included). For PHI, the sensitivity is estimated for each of 6$\times$2 sub-bands, individually spanning a 10\% range in wavelength, under the assumption of $R = 10$. Surface brightness sensitivity is measured per diffraction beam solid angle.
    \par}

\end{table*}

\section{Detector System \label{sec-detector}}

\subsection{Kinetic Inductance Detectors}
\begin{figure*}[ht] 
        \centering
 	\includegraphics[width=\textwidth]{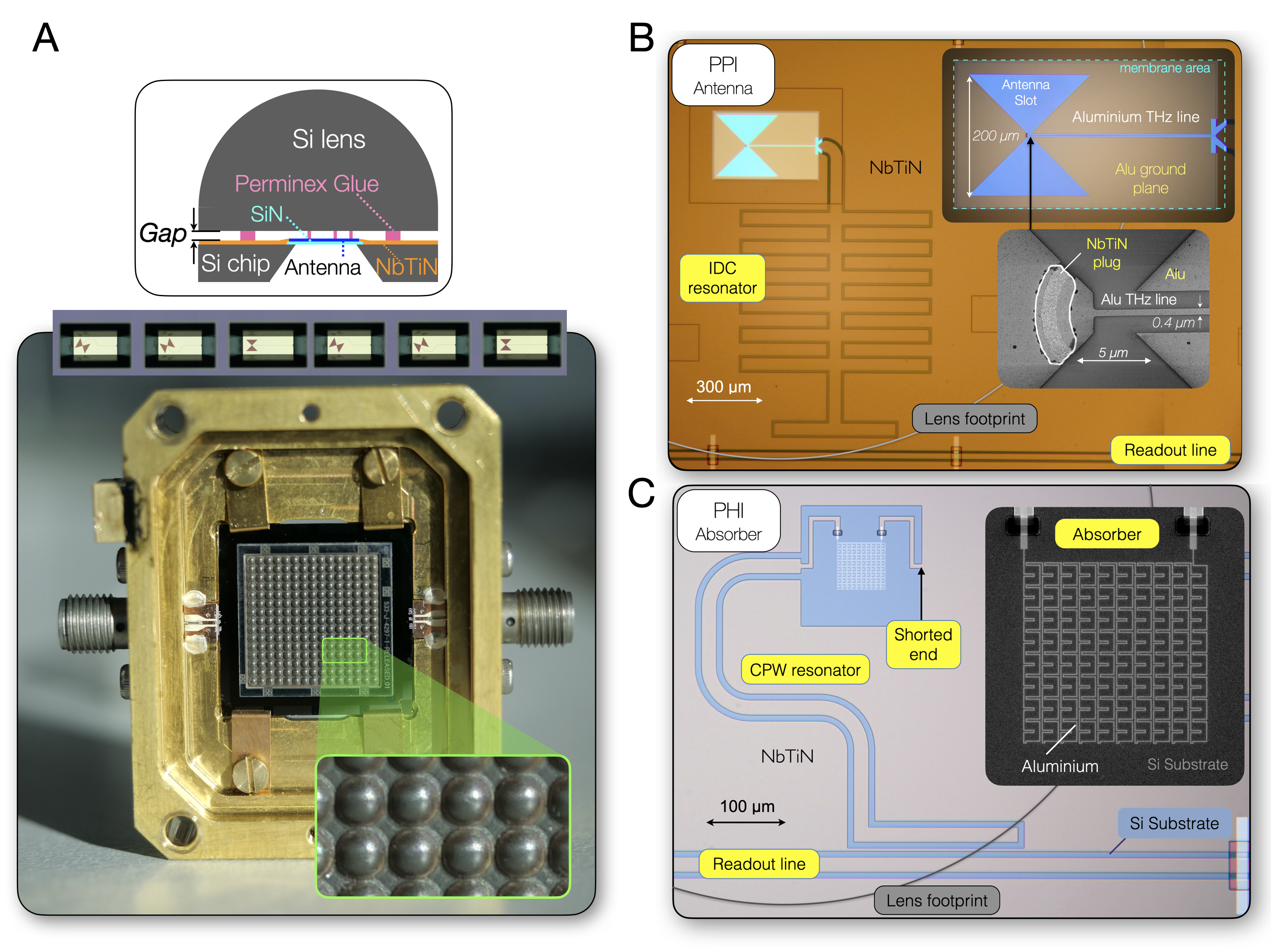}
  	\caption{\label{fig:det}\textbf{Left panel (A):} Photograph of an assembled PPI prototype consisting of a KID array leaky-lens antenna coupled detectors. The lens array is glued using perminex pillars that define a required 4\,$\mu$m gap between lens array and the antenna feeds, fabricated on 100\,nm SiN membranes. The sketch on top shows the cross sectional view of each pixel. The inset shows 6 antennas’, clocked with 120° angles. \textbf{Top right panel (B):} Micrograph of a single detector pixel. \textbf{Bottom right panel (C):} Micrograph of a PRIMAger PHI prototype, using lens-absorber coupling with dual polarization radiation detection.}
\end{figure*}

In the two focal planes of PRIMAger, detector arrays are populated by Microwave Kinetic Inductance detectors, pioneered by Day et al. \,(2001)\cite{day03}.
In the polarimetry focal plane, PPI, we use leaky lens-antenna coupled NbTiN-aluminum devices as presented by Baselmans et al. (2023)\cite{Baselmans:AA23}. 
In these devices, each pixel has a leaky slot antenna on a thin membrane, coupled to a Si lens which is part of a monolithic lens array that is glued to the detector chip using Perminex glue, as shown in Fig.~\ref{fig:det}. 
In the hyperspectral focal plane (PHI), the antenna is replaced by an absorber structure allowing less strict alignment tolerances between the detector and lens array. 
Several prototype arrays have been tested for PRIMA, operating at either 1.5 or 12\,THz. Both the 1.5\,THz (200\,$\mu$m)\cite{Baselmans:AA23} and 12\,THz (24\,$\mu$m)\cite{Day24} prototype devices were measured to be background limited over the entire range of expected powers absorbed in the detectors. 
Additionally, the noise spectra are fully white within this range of powers, with a 1/\textit{f} noise increase below -3dB at 0.1\,Hz as required, and the dark Noise Equivalent Power (NEP) is around 3$\times$10$^{-20}$\,W\,Hz$^{-0.5}$\cite{Baselmans:AA23,Day24}, well below the PRIMAger requirements.  

The KID arrays are housed inside the detector modules, which provide the required thermal, optical and magnetic environment for optimal operation. 
In the case of the PPI, the detector arrays themselves provide polarimetric filtering using polarization sensitive detectors, with the antennas oriented at one of three different angles.
The detector system is currently at TRL5, and during Phase A the development activities will focus on bringing the detector modules to TRL6. 
This will include warm unpowered vibration test of launch loads, radiation harm tests, cosmic ray susceptibility and thermal cycles. 
The performance of the detector assembly (assembled detector module) will be verified before and after the environmental tests happen.

KIDs engineered for the ultra-low background loading of PRIMAger are designed to have a high responsitivity, i.e. a high kinetic inductance fraction and small volume. The detailed designs balance the requirement for background-limited performance at the lowest astrophysical backgrounds with the desire to maintain calibration accuracy for significantly brighter sources and to avoid loss in effective detector yield due to resonator frequency `collisions’ caused by those sources.

\subsection{Detector Readout}

\begin{figure}
\begin{center}
\begin{tabular}{c}
\includegraphics[width=0.9\textwidth]{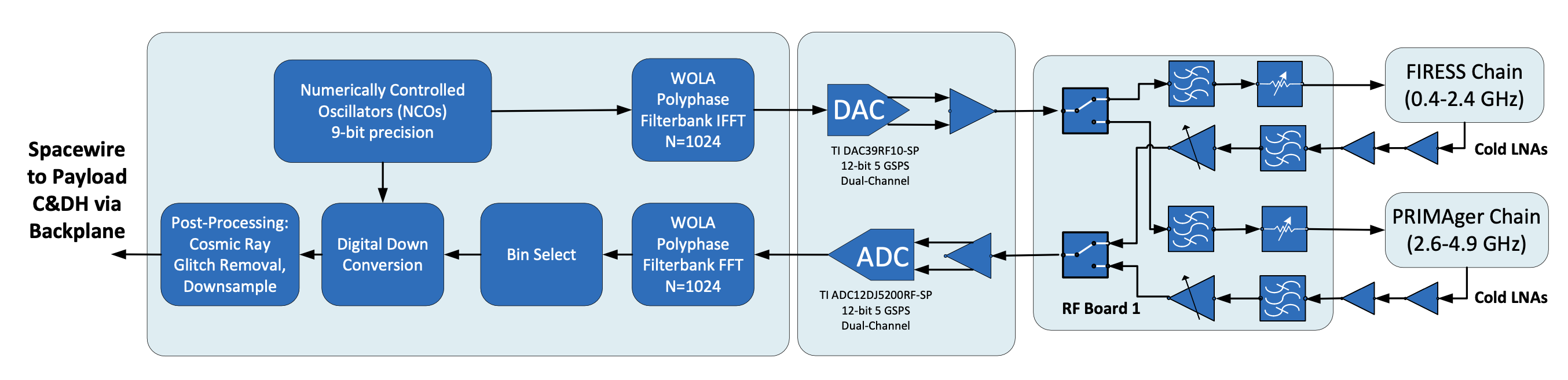}
\end{tabular}
\end{center}
\caption 
{ \label{fig:readout_diagram}
Block diagram of a single PRIMA KID readout chain. The 8 chains are switched between FIRESS and PRIMAger with filtering to allow readout of FIRESS in the first Nyquist zone and PRIMAger in the second Nyquist zone of the ADC and DAC when sampling at 5 Gsps. A high-heritage SpaceCube Mini v3.0 board (labeled ``FPGA Board'') houses a Xilinx Kintex Ultrascale KU060 FPGA and has high-speed interfaces with the ADC/DAC digitizer board. The RF electronics provide switching between the two instruments, filtering, and adjustment of overall gain for system noise optimization.}
\end{figure}

Frequency multiplexing readout electronics for the KID arrays, developed by NASA/GSFC, are shared by FIRESS and PRIMAger via RF switching, with one instrument operating at a time. In each of our eight signal chains, a comb of tones, one for each KID in the array, is synthesized via a weighted overlap-and-add (WOLA) linear time-varying polyphase filterbank (PFB) inverse fast Fourier transform (IFFT) on a Xilinx Kintex KU060 Ultrascale Field-Programmable Gate Array (FPGA). The comb is converted to analog and sent through coaxial cables to the KID arrays. After interacting with the KID array, the modified comb returns through coaxial cables with amplification by low-noise amplifiers (LNAs) at 18 K and 100 K. The returning comb of tones is digitized and processed through a WOLA PFB digital spectrometer, yielding per-detector, time-ordered data. Fig.~\ref{fig:readout_diagram} shows the signal flow in a single readout chain. The algorithms have been used for over a decade including in ground and balloon-borne instruments.~\cite{Gordon_2016_Readout, Sinclair2022_CCAT_Readout}

The Analog-to-Digital Conversion (ADC) and Digital-to-Analog Conversion (DAC) rates of 5 giga-samples per second (Gsps) place the FIRESS readout band of 0.4–2.4 GHz in the first Nyquist zone and the PRIMAger 2.6-4.9\,GHz readout band in the second Nyquist zone. FIRESS and PRIMAger will not operate at the same time, and an RF board provides switching between the two instruments, with filtering for band selection and variable gain to set outgoing tone powers. Modular digital
readout electronics use high-heritage SpaceCube boards.~\cite{Petrick15_SpaceCube, Wilson15_SpaceCube, Brewer20_SpaceCube, Sabogal17_SpaceCube, Perryman21_SpaceCube, Kanekal2019_SpaceCube_MERiT_CeREs} The SpaceCube Mini v3.0 FPGA board and associated power supply cards are build-to-print and currently flying on the STP-H9-SCENIC mission on the ISS~\cite{Geist23_SpaceCube}. PRIMA requires a custom ADC/DAC board, backplane, and enclosure, which are adaptations of existing designs.~\cite{Dubayah20_GEDI, Sun20_GEDI}

\section{Opto-mechanical and thermal design \label{sec-design}}

In this section we outline the pre-Phase-A opto-mechanical-thermal design of the instrument, integrating performance and budget requirements as defined at the current stage of the project.

System-wise the PRIMAger focal plane unit consists of two separate structures: the 1-K opto-mechanical structure that is interfaced to the PRIMA optical bench and contains most of the optical elements of the instrument, and the two nearly identical 125-mK structures that contain the PRIMAger detectors. The latter are enclosed in an outer shell mounted on the 1-K structure to isolate them from the radiation background in the instrument bay (at around 4.5\,K). 

\subsection{Mechanical design}

\begin{figure}[ht] 
 	\includegraphics[width=\textwidth]{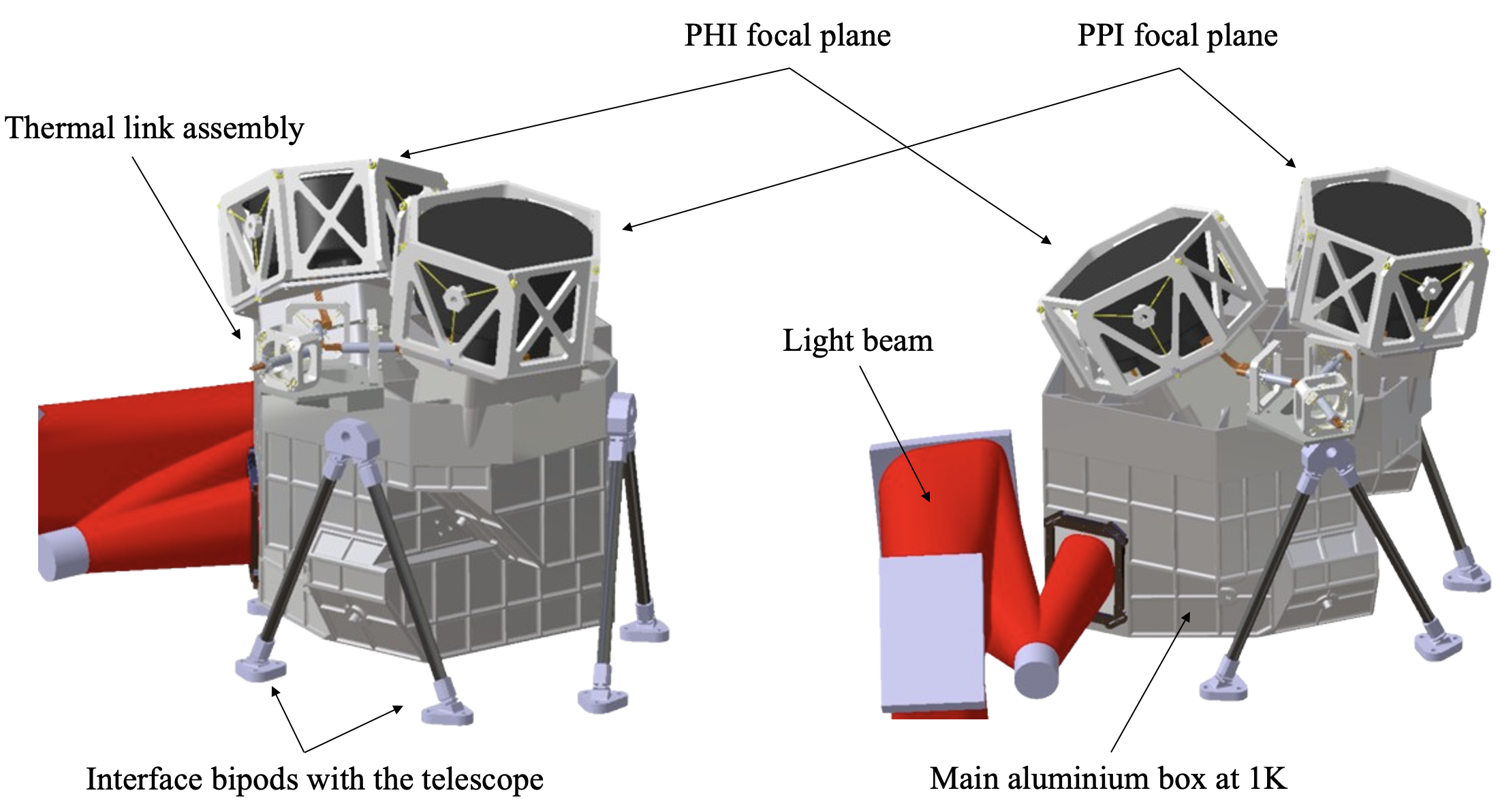}
  	\caption{\label{fig:cad} Opto-mechanical design of PRIMAger viewed from two different angles.} 
\end{figure}

The opto-mechanical design of the PRIMAger instrument features a primary box bolted on a main bench cooled to 1\,K, as illustrated in Fig.~\ref{fig:cad}. These two aluminum structures are light-weighted and reinforced to optimize mass-to-stiffness ratio. The chosen material is aluminum 6061 T6 which is typically used for cryogenic applications. The main box contains aluminum mirrors feeding the two focal plane units that are supported by the main bench. Each mirror is mounted on the box via a mechanical system allowing their accurate positioning, alignment, and stability. The material used to manufacture these mirrors is the same as that of the main structures, ensuring a homothety during cool-down maintaining optical alignment and performance. The two focal plane units, including the detector systems operating at 125\,mK, are bolted to the main bench. The two focal planes are protected by magnetic shielding. Each detector box is supported by kevlar cords system, ensuring a thermal  decoupling between the 1\,K and 125\,mK stages. 
The whole instrument is attached to the PRIMA telescope platform via a kinematic system of three bipods allowing its thermal decoupling from the telescope, particularly the conductive thermal insulation between the 4.5-K telescope platform and the 1-K PRIMAger main box. This opto-mechanical design is compatible with typical space environment conditions:  launch vibration levels of several tens of $g$, followed by cooldown and cryogenic operating temperature.. The total mass of the instrument is $\sim$49\,kg for a volume of 800$\times$600$\times$670\,mm$^3$.

\subsection{Optical design}

\subsubsection{Optical concept}

\begin{figure}[ht] 
        \centering
 	\includegraphics[width=12cm]{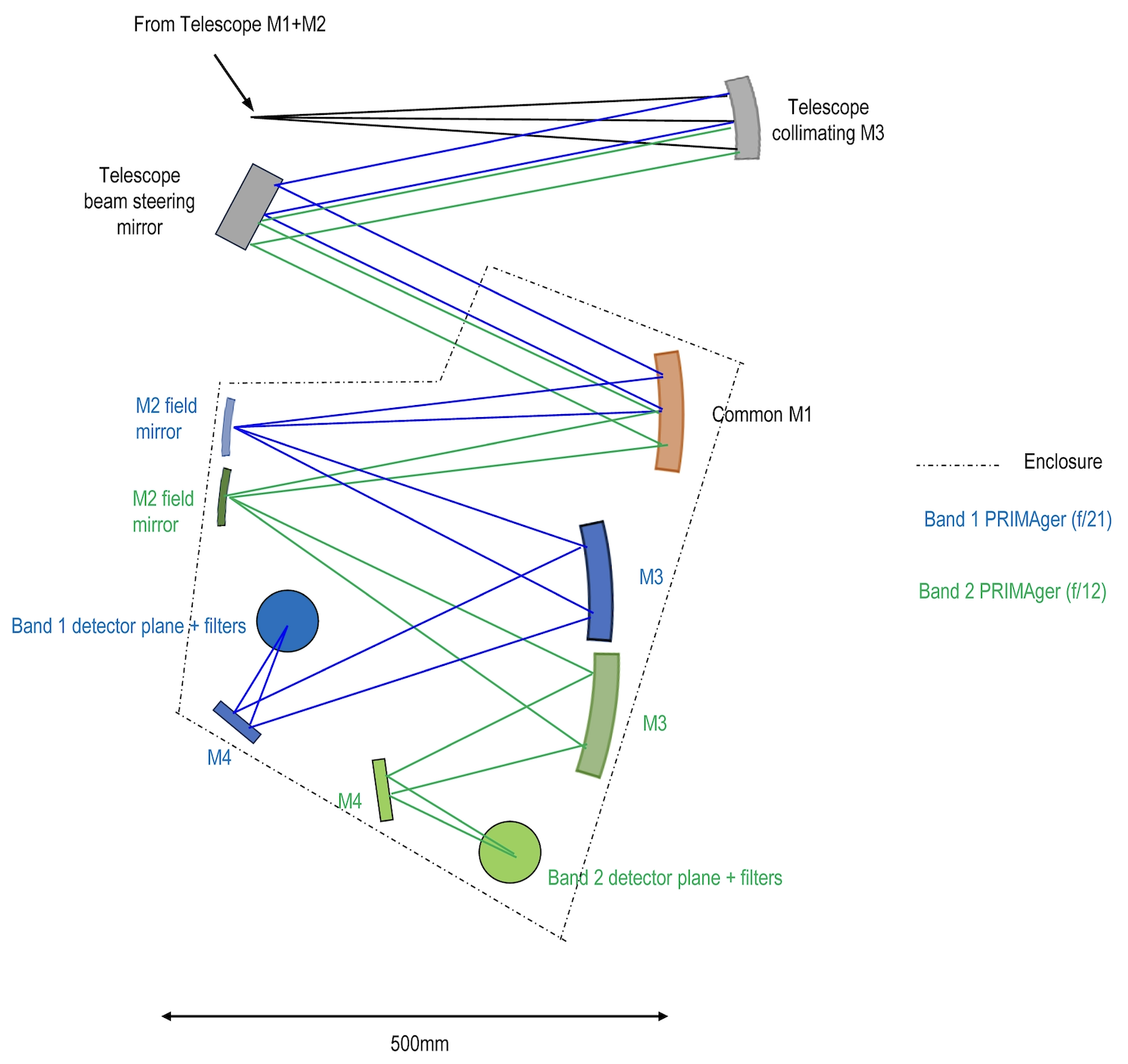}
  	\caption{\label{fig:opt} Optical design of PRIMAger.  The mirrors that are part of the instrument (the dashed line indicates the enclosure perimeter) are colored in blue and green for PHI and PPI, respectively, and in orange for the common M1. Focal plane units are indicated by a circle using the same color code. Gray elements are part of the telescope.}
\end{figure}

The optical layouts of PRIMAger are shown in  Fig.~\ref{fig:opt}. PRIMAger is fed with the 50\,mm collimated optical beam coming from the telescope BSM located at the exit pupil of the telescope, just in front of the PRIMAger box entrance. The BSM also constitutes the entrance pupil of PRIMAger.

Inside the box, a first M1 mirror (common to PHI and PPI beams) forms an intermediate focal plane where PHI and PPI bands are spatially separated. After the common M1 mirror, both bands follow separated light paths with similar optical architectures. M2 field mirrors (with field stops) are placed at this intermediate focal plane and control the size, position, and wandering of the intermediate pupil for PHI and PPI. These intermediate pupils constitute the cold aperture stops for the PRIMAger instrument. They are optimized in size to both maximize throughput and minimize straylight. The 1\,K filters are located at the cold stops.

M3 mirrors forms the image of the intermediate focal plane on the detector planes with the required focal ratios, f/21 and f/12 for PHI and PPI, respectively. M4 mirrors are primarily fold mirrors allowing to fit in the allocated volume, with aspherization in order to optimize image quality.

All mirrors are made of bare aluminum and have freeform surfaces.

Regarding image quality, PHI is diffraction-limited for the whole field of view and for all wavelengths. PPI is diffraction-limited except near the external limit of the field of view. Telecentricity is an important parameter because of the micro-lenses arrays located in front of the detector active surfaces. Telecentricity is $<$0.2° for PHI and $<$0.5° for PPI.

\subsubsection{Linear variable filters}

\begin{figure}[ht] 
        \centering
 	\includegraphics[width=7cm]{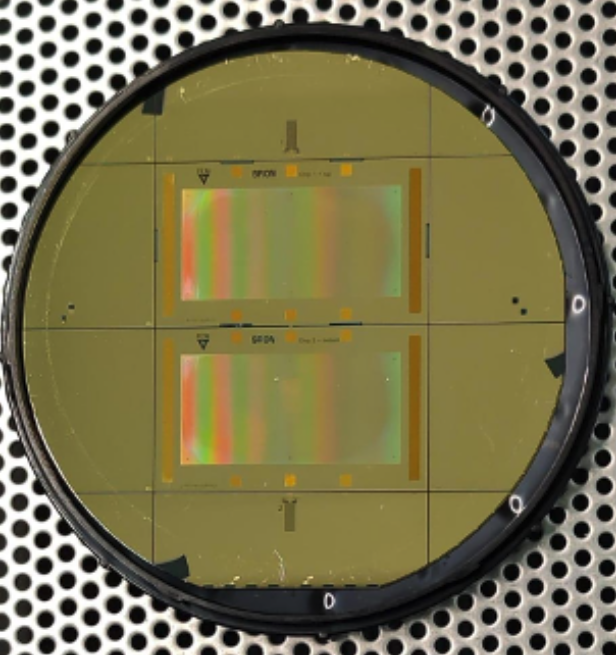}
  	\caption{\label{fig:lvf} PHI1 flight prototype 25–45\,$\mu$m graded filter chips (SRON), demonstrated as dual-layer resonant metal-mesh LVF for PRIMAger.}
\end{figure}

The Linear Variable Filters (LVFs) planned for PRIMAger build upon the established technology of resonant metal-mesh structures (Ade et al. 2006\cite{Ade06}), similar to other filters technologies utilized in PRIMA and described in the following subsection.  
The spectral response of each LVF varies linearly along one axis of the array. Dual-layer scale models at 200 and 375$\mu$m have demonstrated the targeted bandpass characteristics, resolving power, and in-band transmission. Spectral measurements of single-layer prototypes  (Fig.~\ref{fig:lvf}) for the shortwave PRIMAger band (PHI1) show excellent agreement with simulation results.
SRON manufactured a flight-prototype for PHI1 demonstrating graded spectral resolution of approximately R=8 at the short-wavelength limit, while satisfying bandwidth and transmission requirements, for representative sized filters covering the 25–45$\mu$m band. Robustness was demonstrated on a small prototype (20 × 20 mm) with successful vibration and thermal cycling testing.

\subsubsection{Radiation filters}

Optical filtering in PRIMAger, other than the LVFs, is performed using the well-established metal-mesh technology of Cardiff University\cite{Ade06}.  Multiple layers of lithographically patterned Cu meshes are embedded in a low-loss polypropylene matrix.  These quasi-optical devices have been previously deployed in many ground-based and spaceborne facilities including ESA's \textit{Herschel} Observatory\cite{Griffin10,Poglitsch10} and a similar philosophy to their application is used here.  The transmission range is extended to shorter wavelengths in PRIMAger, and the required engineering of the filter designs has been funded by recent ESA, UKSA and STFC awards and demonstrated on NASA's LRO mission\cite{Paige10}.

The PRIMAger filter design has the following elements:  
\begin{itemize}
    \item a low-pass filter mounted on the 1\,K structure at the entrance to the instrument, but exposed to the 4.5\,K ambient radiation.
    \item additional low-pass and high-pass filters at 1\,K.
    \item additional low-pass and high-pass filters at 125\,mK.
\end{itemize}

Low-pass filters deployed at 4.5\,K (in the telescope optics), 1\,K and 125\,mK successively block IR and FIR out-of-band radiation, limiting the power reaching  the detectors.  Combinations of high-pass and low-pass filters define the specific bands of PRIMAger, whilst band-pass filters further define the bandwidths at the array. The combined throughput from the filter chain (excluding the LVF on the PHI) aims to meet a requirement of no worse than 45\%, whilst also reducing out-of-band (IR, optical and UV) radiation to better than 1 part in 10$^{10}$.

\subsection{Thermal design}

The passively-cooled thermal shields of PRIMA are based on the V-groove design of \textit{Spitzer} and \textit{Planck} and achieve a base temperature of 18\,K.
The telescope and instrument interface stages are cooled down to 4.5\,K using a JWST/MIRI-like Joule-Thomson cooler. PRIMAger's opto-mechanical structure is cooled down to 1\,K and incorporates two detector modules cooled down to 125\,mK, thanks to a NASA-Goddard Adiabatic Demagnetization Refrigerator\cite{Dipirro2024}.  This low temperature ADR cooler benefits from Hitomi and XRISM flight heritage. 

The Thermal Link Assembly (TLA) connects the thermal link at the 125\,mK PRIMAger external interface to the two FPAs. This internal thermal link enables to limit the thermal interface to a single point, providing 2.2\,$\mu$W of cooling power, and makes a mechanical damping structure that minimizes the stress on the detector assemblies. At the 1\,K side of this assembly, a system of Kevlar cords, design for 100$g$ static load, supports a titanium mechanical structure that incorporates a high conductive copper bus at 125\,mK.  Flexible copper elements then reach the FPA and minimize mechanical coupling. The overall conductance is close to 1\,$\mu$W/mK. The 1\,K thermal link to the CADR is directly connected to the PRIMAger main bench and provides a cooling power of about 114\,$\mu$W.

\begin{figure}[ht] 
        \centering
 	\includegraphics[width=15cm]{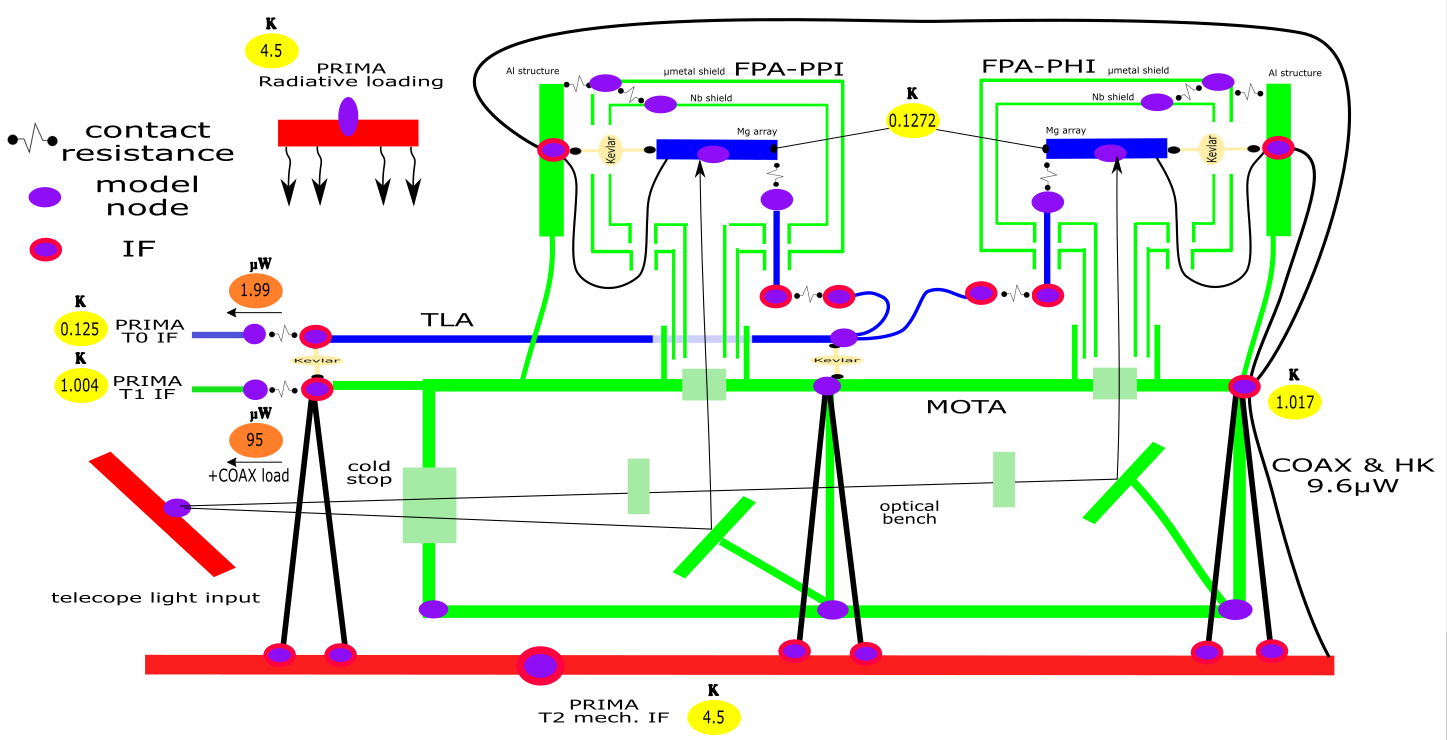}
  	\caption{\label{fig:therm} Simplified view of the thermal architecture and modeling.
The instrument body is made of the Mechanical and Optical Thermal Assembly (MOTA) at 1\,K which incorporates the 125\,mK stage made of two FPAs and the Thermal Link Assembly (TLA). 
 } 
\end{figure}

The thermal modeling, depicted in its early phase on Fig.~\ref{fig:therm}, enables to show as expected low thermal gradient in the opto-mechanical assembly aluminum structure, with less than 20\,mK of temperature gradient. At 125\,mK, the effort has been put on time constants study and thermal mass to be cool down in the first cooling step of the mission. The high thermal conductance of the thermal link (better than 1\,mK/K) provides a time constant of 5\,s, leading to extremely low temperature gradient in the cold configuration.

The PRIMA cryochain does not include redundancy, and the same cooling system will serve both PRIMAger and FIRESS.  The PRIMA team is in charge of the cryochain and thermal links reaching PRIMAger. 
The two instruments do not operate at the same time but will be cold at all times.

\section{Summary  \label{sec-sum}}
In this article, we presented the PRIMAger instrument in its design study phase.
Strongly endorsed by the community\cite{Moullet23}, PRIMAger is a key instrument aboard the PRobe far-Infrared Mission for Astrophysics (PRIMA). It is a highly capable FIR hyperspectral and polarimetric imager designed to address fundamental astrophysical questions and expand observational capabilities in the FIR domain. The instrument consists of two focal planes: the Hyperspectral Imager (PHI) covering the 24–84\,$\mu$m range with a spectral resolution of $R \approx 8$ and the Polarimetric Imager (PPI), operating in four broadband filters spanning 80–264\,$\mu$m, with sensitivity to linear polarization. Both focal planes use state-of-the-art kinetic inductance detectors (KIDs), which ensure exceptional sensitivity and performance.
PRIMAger employs a beam steering mirror and spacecraft scanning to efficiently map large sky areas.
It provides exceptional mapping speeds, angular resolution, and polarimetric sensitivity, enabling studies of dust emission, star formation processes, galactic magnetic fields, and extragalactic evolution. 
PRIMAger’s performance eclipses that of previous FIR missions like \textit{Herschel}/SPIRE, making it a cornerstone for next-generation FIR astronomy and opening vast discovery potential for the international astrophysics community.
A companion paper\cite{Burgarella25} details science cases specific to PRIMAger, highlighting the wealth of science that this instrument will enable.

\section*{Code and data}
This paper does not rest on code or data that would be appropriate to share.

\section*{Disclosures}
The authors declare that there are no financial interests, commercial affiliations, or other potential conflicts of interest that could have influenced the objectivity of this research or the writing of this paper.

\acknowledgments     
 
LC acknowledges support from the French government under the France 2030 investment plan, as part of the Initiative d’Excellence d’Aix-Marseille Université – A*MIDEX AMX-22-RE-AB-101.
LC, MS, and DB acknowledge funding support from CNES.
CET and MJG acknowledge funding support from the UK Space Agency under grant ST/Y005465/1.
This research was carried out in part at the Jet Propulsion Laboratory, California Institute of Technology, under a contract with the National Aeronautics and Space Administration (80NM0018D0004).

\bibliography{report}   

\begin{thebibliography}{10}

\bibitem{Astro2020}
{National Academies of Sciences, Engineering, and Medicine}, {\em {Pathways to Discovery in Astronomy and Astrophysics for the 2020s}}, The National Academies Press, Washington, DC  (2023).

\bibitem{apexcall}
{NASA}, ``Apex call.'' \url{https://explorers.larc.nasa.gov/2023APPROBE/pdf_files/NNH23ZDA021OSummary.pdf}.

\bibitem{Glenn25}
J.~{Glenn}, M.~{Meixner}, C.~M. {Bradford}, {\em et~al.}, ``{The PRIMA mission concept},'' {\em Journal of Astronomical Telescopes, Instruments, and Systems} {\bf 11}, 031628  (2025).

\bibitem{Bradford25}
C.~M. {Bradford}, A.~J. {Kogut}, D.~{Fixsen}, {\em et~al.}, ``{The Far-Infrared Enhanced Survey Spectrometer (FIRESS) for PRIMA: Approach and Estimated Performance},'' {\em Journal of Astronomical Telescopes, Instruments, and Systems} {\bf 11}, 031627  (2025).

\bibitem{Moullet23}
A.~{Moullet}, T.~{Kataria}, D.~{Lis}, {\em et~al.}, ``{PRIMA General Observer Science Book},'' {\em arXiv e-prints} , arXiv:2310.20572  (2023).

\bibitem{Dowell2024}
C.~D. {Dowell}, B.~S. {Hensley}, and M.~{Sauvage}, ``{Simulation of the Far-Infrared Polarimetry Approach Envisioned for the PRIMA Mission},'' {\em arXiv e-prints} , arXiv:2404.17050  (2024).

\bibitem{Waskett07}
T.~J. {Waskett}, B.~{Sibthorpe}, M.~J. {Griffin}, {\em et~al.}, ``{Determining the optimum scan map strategy for Herschel-SPIRE using the SPIRE photometer simulator},'' {\em \mnras} {\bf 381}, 1583--1590  (2007).

\bibitem{Dowell2010}
C.~D. {Dowell}, M.~{Pohlen}, C.~{Pearson}, {\em et~al.}, ``{Status of the SPIRE photometer data processing pipelines during the early phases of the Herschel Mission},'' in {\em Space Telescopes and Instrumentation 2010: Optical, Infrared, and Millimeter Wave},  J.~M. {Oschmann}, Jr., M.~C. {Clampin}, and H.~A. {MacEwen}, Eds., {\em Society of Photo-Optical Instrumentation Engineers (SPIE) Conference Series} {\bf 7731}, 773136  (2010).

\bibitem{Roussel2013}
H.~{Roussel}, ``{Scanamorphos: A Map-making Software for Herschel and Similar Scanning Bolometer Arrays},'' {\em \pasp} {\bf 125}, 1126  (2013).

\bibitem{Krause06}
O.~{Krause}, D.~{Lemke}, R.~{Hofferbert}, {\em et~al.}, ``{The cold focal plane chopper of HERSCHEL's PACS instrument},'' in {\em Optomechanical Technologies for Astronomy},  E.~{Atad-Ettedgui}, J.~{Antebi}, and D.~{Lemke}, Eds., {\em Society of Photo-Optical Instrumentation Engineers (SPIE) Conference Series} {\bf 6273}, 627325  (2006).

\bibitem{Griffin10}
M.~J. {Griffin}, A.~{Abergel}, A.~{Abreu}, {\em et~al.}, ``{The Herschel-SPIRE instrument and its in-flight performance},'' {\em \aap} {\bf 518}, L3  (2010).

\bibitem{Burgarella25}
D.~{Burgarella}, M.~{B\'ethermin}, A.~{Boselli}, {\em et~al.}, ``{PRIMAger General Observer programs: a pi-sr Infrared Survey and other wide-field programs},'' {\em Journal of Astronomical Telescopes, Instruments, and Systems} {\bf 11}, 031639  (2025).

\bibitem{Bethermin24}
M.~{B\'ethermin}, A.~D. {Bolatto}, F.~{Boulanger}, {\em et~al.}, ``{Confusion of extragalactic sources in the far infrared: a baseline assessment of the performance of PRIMAger in intensity and polarization},'' {\em \aap}   (2024).

\bibitem{Donnellan24}
J.~M.~S. {Donnellan}, S.~J. {Oliver}, M.~{B{\'e}thermin}, {\em et~al.}, ``{Overcoming confusion noise with hyperspectral imaging from PRIMAger},'' {\em \mnras} {\bf 532}, 1966--1979  (2024).

\bibitem{Poglitsch10}
A.~{Poglitsch}, C.~{Waelkens}, N.~{Geis}, {\em et~al.}, ``{The Photodetector Array Camera and Spectrometer (PACS) on the Herschel Space Observatory},'' {\em \aap} {\bf 518}, L2  (2010).

\bibitem{day03}
P.~Day, H.~LeDuc, B.~Mazin, {\em et~al.}, ``A broadband superconducting detector suitable for use in large arrays,'' {\em Nature} {\bf 425}, 817--821  (2003).

\bibitem{Baselmans:AA23}
J.~J.~A. {Baselmans}, F.~{Facchin}, A.~{Pascual Laguna}, {\em et~al.}, ``{Ultra-sensitive THz microwave kinetic inductance detectors for future space telescopes},'' {\em \aap} {\bf 665}, A17  (2022).

\bibitem{Day24}
P.~K. {Day}, N.~F. {Cothard}, C.~{Albert}, {\em et~al.}, ``{A 25-micron single photon sensitive kinetic inductance detector},'' {\em arXiv e-prints} , arXiv:2404.10246  (2024).

\bibitem{Gordon_2016_Readout}
S.~Gordon, B.~Dober, A.~Sinclair, {\em et~al.}, ``An open source, fpga-based lekid readout for blast-tng: Pre-flight results,'' {\em Journal of Astronomical Instrumentation} {\bf 05}  (2016).

\bibitem{Sinclair2022_CCAT_Readout}
A.~K. Sinclair, R.~C. Stephenson, C.~A. Roberson, {\em et~al.}, ``{CCAT-prime: RFSoC based readout for frequency multiplexed kinetic inductance detectors},'' in {\em Millimeter, Submillimeter, and Far-Infrared Detectors and Instrumentation for Astronomy XI},  J.~Zmuidzinas and J.-R. Gao, Eds.,  {\bf 12190}, 121900W, International Society for Optics and Photonics, SPIE  (2022).

\bibitem{Petrick15_SpaceCube}
D.~{Petrick}, N.~{Gill}, M.~{Hassouneh}, {\em et~al.}, ``Adapting the {SpaceCube} v2.0 data processing system for mission-unique application requirements,'' in {\em IEEE Aerospace Conference},   (2015).

\bibitem{Wilson15_SpaceCube}
C.~{Wilson}, J.~{Stewart}, P.~{Gauvin}, {\em et~al.}, ``{CSP} hybrid space computing for {STP-H5/ISEM} on {ISS},'' in {\em 19th Annu. AIAA/USU Conf. on Small Satellites},  {\em SSC15-III-10}  (2015).

\bibitem{Brewer20_SpaceCube}
C.~{Brewer}, N.~{Franconi}, R.~{Ripley}, {\em et~al.}, ``{NASA} {SpaceCube} intelligent multi-purpose system for enabling remote sensing, communication, and navigation in mission architectures,'' in {\em 34th Annu. AIAA/USU Conf. on Small Satellites},  {\em SSC20-VI-07}  (2020).

\bibitem{Sabogal17_SpaceCube}
S.~{Sabogal}, P.~{Gauvin}, B.~{Shea}, {\em et~al.}, ``Spacecraft supercomputing experiment for {STP-H6},'' in {\em 31st Annu. AIAA/USU Conf. on Small Satellites},  {\em SSC17-XIII-02}  (2017).

\bibitem{Perryman21_SpaceCube}
N.~{Perryman}, T.~{Schwarz}, T.~{Cooke}, {\em et~al.}, ``{STP-H7-CASPR}: A transition from mission concept to launch,'' in {\em 35th Annu. AIAA/USU Conf. on Small Satellites},  {\em SSC21-WKII-08}  (2021).

\bibitem{Kanekal2019_SpaceCube_MERiT_CeREs}
S.~G. Kanekal, L.~Blum, E.~R. Christian, {\em et~al.}, ``The merit onboard the ceres: A novel instrument to study energetic particles in the earth's radiation belts,'' {\em Journal of Geophysical Research: Space Physics} {\bf 124}(7), 5734--5760  (2019).

\bibitem{Geist23_SpaceCube}
A.~{Geist}, G.~{Crum}, C.~{Brewer}, {\em et~al.}, ``{NASA} {SpaceCube} next-generation artificial-intelligence computing for {STP-H9-SCENIC} on {ISS},'' in {\em 37th Annu. AIAA/USU Conf. on Small Satellites},  {\em SSC23-P1-32}  (2023).

\bibitem{Dubayah20_GEDI}
R.~Dubayah, J.~B. Blair, S.~Goetz, {\em et~al.}, ``The global ecosystem dynamics investigation: High-resolution laser ranging of the earth’s forests and topography,'' {\em Science of Remote Sensing} {\bf 1}, 100002  (2020).

\bibitem{Sun20_GEDI}
X.~Sun, J.~B. Blair, J.~L. Bufton, {\em et~al.}, ``{Advanced silicon avalanche photodiodes on NASA's Global Ecosystem Dynamics Investigation (GEDI) mission},'' in {\em Photonic Instrumentation Engineering VII},  Y.~Soskind and L.~E. Busse, Eds.,  {\bf 11287}, 1128713, International Society for Optics and Photonics, SPIE  (2020).

\bibitem{Ade06}
P.~A.~R. {Ade}, G.~{Pisano}, C.~{Tucker}, {\em et~al.}, ``{A review of metal mesh filters},'' in {\em Millimeter and Submillimeter Detectors and Instrumentation for Astronomy III},  J.~{Zmuidzinas}, W.~S. {Holland}, S.~{Withington}, {\em et~al.}, Eds., {\em Society of Photo-Optical Instrumentation Engineers (SPIE) Conference Series} {\bf 6275}, 62750U  (2006).

\bibitem{Paige10}
D.~A. {Paige}, M.~C. {Foote}, B.~T. {Greenhagen}, {\em et~al.}, ``{The Lunar Reconnaissance Orbiter Diviner Lunar Radiometer Experiment},'' {\em \ssr} {\bf 150}, 125--160  (2010).

\bibitem{Dipirro2024}
M.~{DiPirro}, P.~{Shirron}, A.~{Jahromi}, {\em et~al.}, ``{The continuous adiabatic demagnetization refrigerator for the probe far-infrared mission for astrophyiscs (PRIMA)},'' in {\em Space Telescopes and Instrumentation 2024: Optical, Infrared, and Millimeter Wave},  L.~E. {Coyle}, S.~{Matsuura}, and M.~D. {Perrin}, Eds., {\em Society of Photo-Optical Instrumentation Engineers (SPIE) Conference Series} {\bf 13092}, 130923A  (2024).

\end{thebibliography}
\bibliographystyle{spiejour}   

\listoffigures
\listoftables

\end{spacing}
\end{document}